\documentclass[conference]{IEEEtran}

\DeclareRobustCommand{\IEEEauthorrefmark}[1]{\smash{\textsuperscript{\footnotesize #1}}}

\usepackage{xspace}
\usepackage{amsmath,amsfonts}
\usepackage{algorithmic}
\usepackage{algorithm}
\usepackage{array}

\usepackage{textcomp}
\usepackage{stfloats}
\usepackage{url}
\usepackage{verbatim}
\usepackage{graphicx}
\usepackage{cite}
\usepackage{multirow}
\usepackage{xcolor}
\usepackage{ulem} 
\usepackage[hidelinks,bookmarks=false]{hyperref} 
\usepackage{orcidlink}
\usepackage{lipsum}

\usepackage{tikz}
\usepackage{subcaption}
\graphicspath{{figs/}}

\newcommand{\scum}          {SC$\mu$M\xspace}

\begin{document}

\title{LightCal: Lightweight Optical-Pulse Bootstrap Calibration for Crystal-Free BLE Radios}

\author{
    \IEEEauthorblockN{
        Cheng Wang\IEEEauthorrefmark{1},
        Titan Yuan\IEEEauthorrefmark{2},        
        David Burnett\IEEEauthorrefmark{3},
        Filip Maksimovic\IEEEauthorrefmark{4},
        Kristofer S.J. Pister\IEEEauthorrefmark{2},
        Tengfei Chang\IEEEauthorrefmark{1}
    }

    \IEEEauthorblockA{\IEEEauthorrefmark{1}\textit{HKUST (Guangzhou)}, Guangzhou, China
    cwang199@connect.hkust-gz.edu.cn, tengfeichang@hkust-gz.edu.cn}

    \IEEEauthorblockA{\IEEEauthorrefmark{2}\textit{UC Berkeley}, Berkeley, CA, USA
    \{titan, ksjp\}@berkeley.edu}

    \IEEEauthorblockA{\IEEEauthorrefmark{3}\textit{Villanova University}, Villanova, PA, USA
    david.burnett@villanova.edu}

    \IEEEauthorblockA{\IEEEauthorrefmark{4}\textit{Inria}, Paris, France
    filip.maksimovic@inria.fr}
}

\maketitle

\begin{abstract}

Crystal-free Bluetooth Low Energy (BLE) radios remove the off-chip high-frequency crystal oscillator and can therefore reduce the cost, size, and integration complexity of Internet of Things (IoT) nodes. However, they face a fundamental bootstrap problem: before a node can communicate over RF, it must first obtain a sufficiently accurate carrier-frequency reference. Existing approaches typically rely on RF beacons, already-connected nodes, or search-based channel acquisition, which can incur long startup latency and provide limited feedback when the initial carrier offset is large.
This paper presents \textit{LightCal}, a lightweight bootstrap calibration method that uses periodic optical pulses as an external timing reference for crystal-free BLE radios. LightCal is designed for highly resource-constrained platforms and requires only simple optical pulse reception. We implement LightCal on \scum, a crystal-free IoT platform and use a commercial HTC Lighthouse V1 base station as an unmodified off-the-shelf optical pulse source. Experimental results show that pulse accumulation substantially improves the effective timing stability of Lighthouse sync pulses on SC$\mu$M and enables practical BLE bootstrap calibration. In the current \scum prototype, optical calibration brings the RF carrier into a bounded residual-error range, and the remaining offset is resolved by a narrow transmit-time fine sweep. The results demonstrate that optical pulse references can provide a practical pre-RF bootstrap calibration path for crystal-free and highly integrated IoT platforms.

\end{abstract}



\section{Introduction}

The continued growth of Internet of Things (IoT) deployments is pushing node design toward smaller size, lower cost, lower power, and higher integration. In this context, crystal-free radios are attractive because eliminating the off-chip high-frequency crystal oscillator can reduce cost, packaging complexity, and power consumption \cite{maksimovic2019crystal,song202030,alghaihab202030}. However, this also creates a fundamental bootstrap problem: before a crystal-free BLE node can communicate reliably, it must first obtain a sufficiently accurate carrier-frequency reference \cite{wheeler2017crystal}.

Existing methods typically derive the initial frequency reference from RF packets, search-based channel acquisition, or already-connected nodes \cite{chang2020quickcal,song202030}. Recent work \cite{yuan2026automatic}, has further integrated initial calibration into the self-joining process of crystal-free radios using network-side packet exchanges and acknowledgments. However, these approaches still rely on RF-side search and opportunistic feedback; during the earliest bootstrap stage, when the initial carrier offset is large, such feedback may remain sparse or unavailable.

\begin{figure}[!t]
	\centering
	\includegraphics[width=\columnwidth]{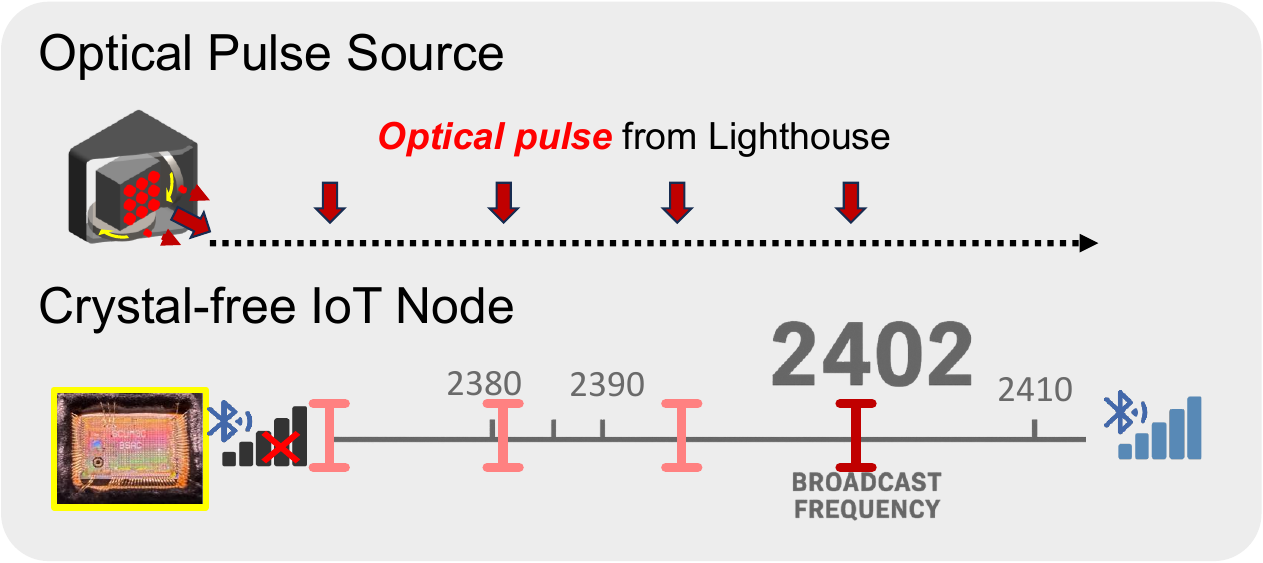}
	\caption{Overview of the LightCal principle.}
	\label{fig:overview}
\end{figure}

In this paper, we show that periodic optical pulses can serve as a practical pre-RF timing reference for bootstrap carrier calibration. A node only needs to passively receive the pulses, while one optical source can provide a shared reference to many nodes. This approach is especially appealing for resource-constrained platforms because optical reception can be simple, and \scum already integrates an on-chip optical receiver \cite{maksimovic2019crystal,wheeler2019low}.

Based on this observation, we propose \textit{LightCal}, a lightweight optical-pulse bootstrap calibration method for crystal-free BLE radios (Fig.~\ref{fig:overview}). LightCal uses periodic optical pulses to define calibration windows and tunes the RF carrier using LC-counter measurements together with lightweight filtering logic. We implement LightCal on \scum and validate it using the platform's on-chip optical receiver together with an unmodified HTC Lighthouse V1 base station as the optical pulse source. Experimental results show that pulse accumulation substantially improves the effective timing stability of Lighthouse sync pulses on SC$\mu$M and enables practical BLE bootstrap calibration. In the current \scum prototype, optical calibration brings the RF carrier into a bounded residual-error range, and the remaining offset is resolved by a narrow transmit-time fine sweep.

The main contributions of this paper are as follows:
\begin{enumerate}
    \item We identify pre-RF carrier bootstrap as a key challenge for crystal-free BLE and show that periodic optical pulses provide a practical external reference.
    \item We design a lightweight calibration pipeline based on pulse accumulation, LC-counter comparison, and differential filtering, making the method suitable for highly resource-constrained devices.
    \item We implement LightCal on \scum and validate it with an unmodified commercial Lighthouse V1 source, showing that it enables practical BLE advertising and can coexist with Lighthouse-based localization.
\end{enumerate}

\section{Motivation and Related Work}

Crystal-free IoT nodes do not include a high-precision external crystal oscillator and must therefore derive an RF frequency reference from other sources. To meet the carrier-frequency accuracy required by wireless communication, a crystal-free node typically keeps adjusting its internal LC oscillator until it can successfully receive a packet or otherwise infer a usable channel setting. Although this process can be accelerated by deploying broadcasters at multiple frequency points \cite{chang2020quickcal}, the initial procedure remains largely open loop when no RF feedback is yet available \cite{wheeler2017crystal}.

Most prior calibration methods focus on frequency maintenance after a node has already joined the network \cite{stanislowski2013adaptive,alghaihab202030,ding20200,xu2016energy,movassaghi2014wireless}. By periodically exchanging packets or timestamps, these methods can maintain synchronization and frequency accuracy as long as connectivity is preserved \cite{chang2021surviving}. However, they do not directly address bootstrap calibration before any successful RF communication is available.

Compared with blind RF scanning, adding a second sensing modality can improve bootstrap calibration. Under the hardware constraints of typical IoT devices, optical-pulse-based calibration is attractive because the receiver can be simple and passive. In addition, one optical source can provide a shared reference to many nodes within line of sight without occupying RF channels. This is especially relevant for resource-constrained platforms such as \scum, which already integrates an on-chip optical receiver \cite{wheeler2019low}.
Recent network-based approaches have begun to address bootstrap calibration during self-joining, but they remain RF-driven and protocol-dependent\cite{yuan2026automatic}.

To our knowledge, prior work has not used periodic optical pulses as a practical pre-RF carrier-calibration reference for crystal-free BLE radios. LightCal fills this gap by combining a simple optical reference with lightweight node-side processing that fits the constraints of \scum-class platforms.

\section{Hardware Platform}

\begin{figure}[t]
    \centering
    \includegraphics[width=0.9\linewidth]{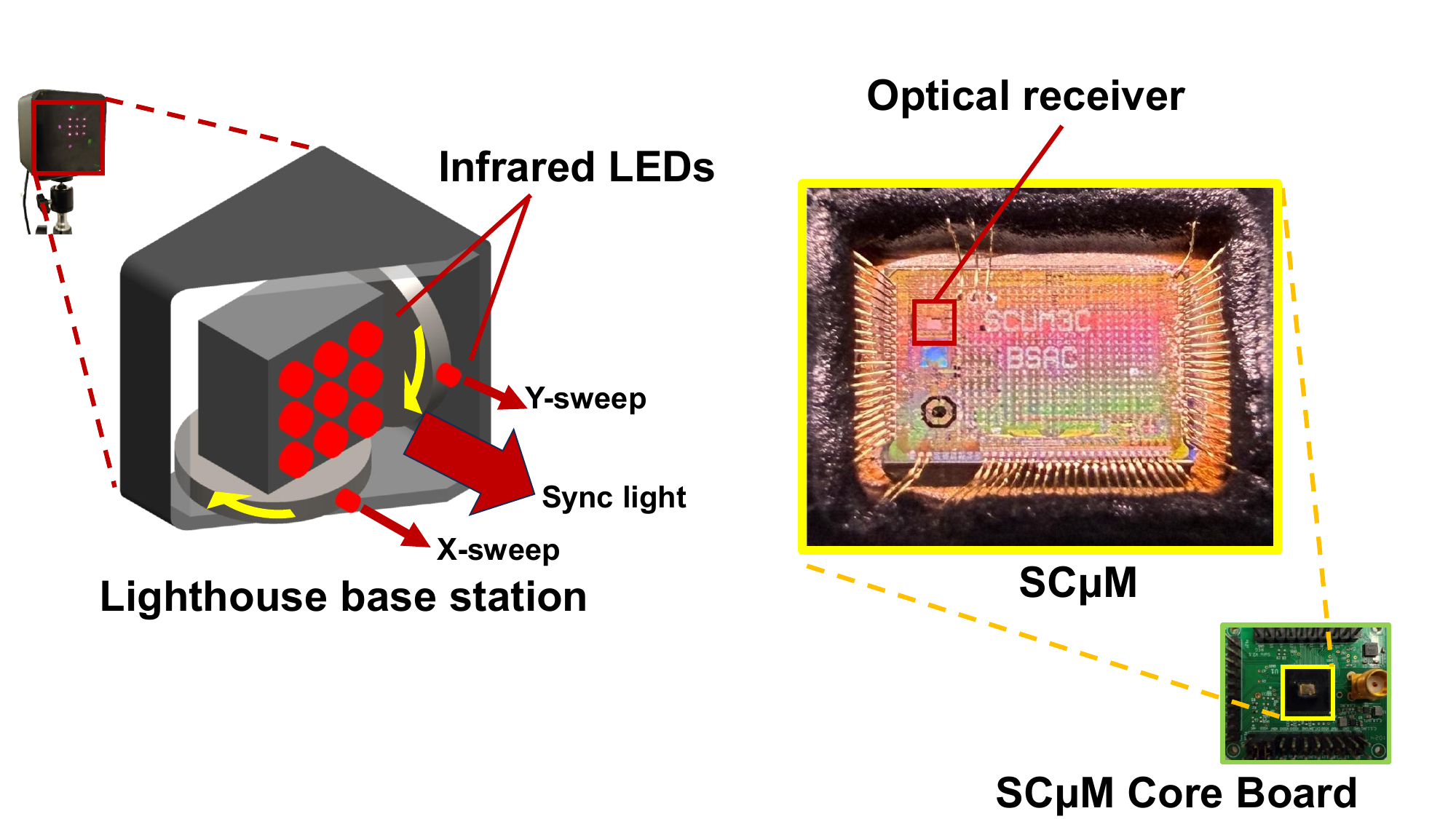}
    \caption{SC$\mu$M and the HTC Lighthouse V1 optical pulse source.}
    \label{fig:hardware}
\end{figure}

\subsection{Crystal-Free BLE Radio: SC$\mu$M}

SC$\mu$M is the crystal-free platform used to implement and validate LightCal (Fig.~\ref{fig:hardware}). It integrates a Cortex-M0 MCU operating at a default frequency of 5 MHz and uses only on-chip oscillators, without any off-chip crystal reference. In particular, SCµM uses an LC oscillator to generate the 2.4 GHz RF carrier, while its other clocks are derived from on-chip RC oscillators \cite{maksimovic2019crystal}. SC$\mu$M also includes an on-chip optical receiver for infrared pulse detection and wireless optical programming \cite{wheeler2019low}. A functional SC$\mu$M-based BLE node requires only an external 1.8~V power supply and no off-chip timing components. \scum is also designed to be compatible with both BLE and IEEE 802.15.4, making it a representative platform for studying crystal-free wireless communication \cite{maksimovic2019crystal}.

\subsection{Lighthouse Base Station}

HTC Lighthouse V1 is a commercial infrared virtual-reality tracking system that emits periodic synchronization pulses followed by mechanical optical sweeps. The optical receiver on SC$\mu$M can detect these pulses and has also been used for Lighthouse-based localization \cite{kilberg2020accurate}. The synchronization pulse width is typically 60--130~$\mu$s, the sweep pulse is on the order of 10~$\mu$s, and the nominal synchronization rate is 120~Hz. Because the sweep is mechanical, the synchronization period fluctuates around its nominal value.

\section{LightCal Design}

LightCal targets the pre-RF bootstrap stage of a crystal-free BLE node. The goal is not to recover an exact carrier frequency in one step, but to bring the RF carrier close enough to a BLE channel that packet transmission becomes feasible. In our \scum implementation, this design must tolerate two practical constraints: Lighthouse sync pulses exhibit non-negligible short-term jitter, and \scum can observe optical events only through software polling rather than hardware interrupts. LightCal addresses these constraints through windowed pulse accumulation, differential filtering, and a final fine-grained transmit-time sweep.

\subsection{Optical Pulses as a Frequency Reference}

\begin{figure*}[t]
  \centering
  \begin{subfigure}[t]{0.23\linewidth}
    \centering
    \includegraphics[width=\linewidth]{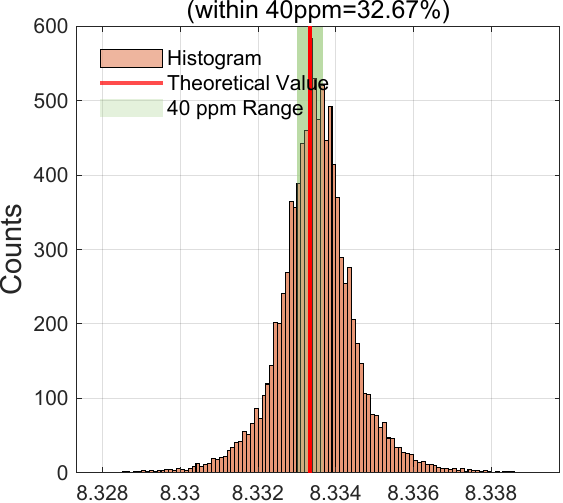}
    \caption{Raw optical-receiver periods}
    \label{fig:period_opt_origin}
  \end{subfigure}
  \begin{subfigure}[t]{0.23\linewidth}
    \centering
    \includegraphics[width=\linewidth]{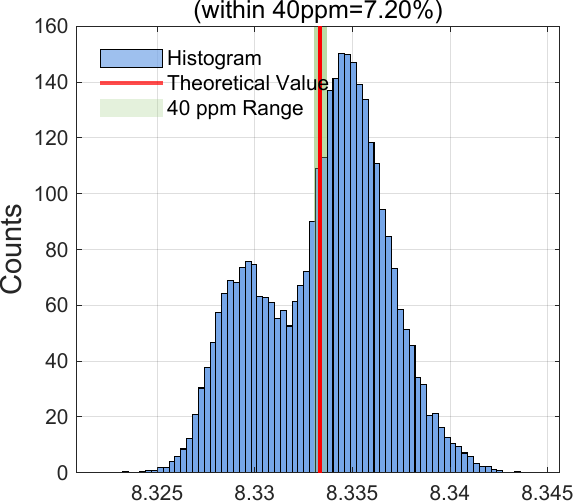}
    \caption{Raw software-polled periods}
    \label{fig:period_soft_origin}
  \end{subfigure}
  \begin{subfigure}[t]{0.23\linewidth}
    \centering
    \includegraphics[width=\linewidth]{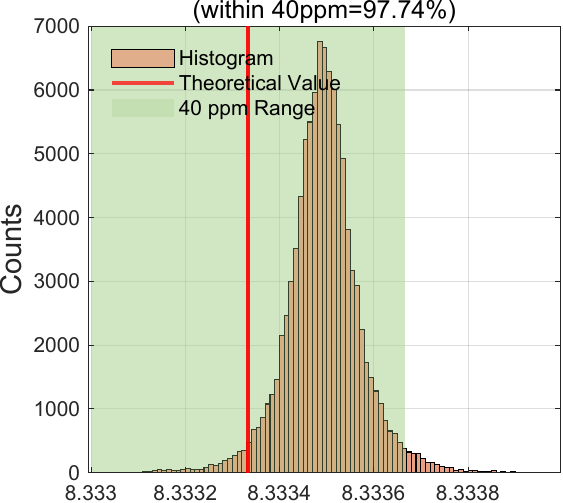}
    \caption{Accumulated optical-receiver\\ periods}
    \label{fig:period_opt_filter}
  \end{subfigure}
  \begin{subfigure}[t]{0.23\linewidth}
    \centering
    \includegraphics[width=\linewidth]{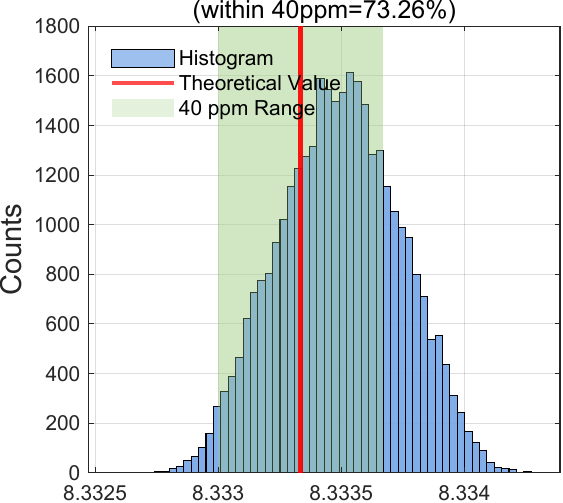}
    \caption{Accumulated software-polled\\ periods}
    \label{fig:period_soft_filter}
  \end{subfigure}

  \caption{Distribution of Lighthouse synchronization-pulse periods measured on SC$\mu$M. Accumulating 12 consecutive sync periods substantially improves effective timing stability.}
  \label{fig:period_distribution}
\end{figure*}

Although Lighthouse uses relatively precise motors, its synchronization timing still contains jitter. The nominal synchronization frequency is 120~Hz, but the actual period fluctuates because of sweep-motor speed variation. Prior work has shown that such jitter can be reduced substantially by accumulating multiple periods \cite{wang2024simultaneous}. In our setting, an additional challenge is that SC$\mu$M does not provide a hardware interrupt for the optical receiver, so pulse timing must be obtained through software polling.

To quantify the impact of both Lighthouse jitter and software polling, we sampled Lighthouse pulses for more than ten minutes. We measured the optical-receiver signal pin of SC$\mu$M directly to obtain the hardware-level period distribution, and we also recorded the pin level through software polling to obtain the periods actually used by LightCal. Fig.~\ref{fig:period_opt_origin} shows that the hardware-level synchronization periods are approximately Gaussian, whereas software polling introduces additional modes in the measured distribution (Fig.~\ref{fig:period_soft_origin}) 
This behavior is likely due to the discrete sampling granularity and execution-time jitter of the polling loop, which quantize pulse-detection timing and occasionally shift the inferred pulse boundaries.
In raw form, neither distribution satisfies the BLE carrier-frequency tolerance of 40~ppm with sufficiently high reliability.

To improve timing stability, we apply accumulation-based filtering. Specifically, we form one effective measurement from every 12 consecutive synchronization pulses, corresponding to a window of about 100~ms. This choice reflects a practical tradeoff between timing stability and calibration latency on SC$\mu$M. Accumulation significantly increases the fraction of measurements that fall within the 40~ppm equivalent range. For the hardware-level signal, the in-spec ratio increases from 32.67\% to 97.74\% (Fig.~\ref{fig:period_opt_filter}). For the software-polled signal actually used by LightCal, the in-spec ratio increases from 7.20\% to 73.26\% (Fig.~\ref{fig:period_soft_filter}). These results show that although polling-induced uncertainty remains significant, a short accumulation window is sufficient to make optical timing practically useful for bootstrap calibration.

\subsection{Tuning Principle}

SC$\mu$M generates the RF carrier from an LC oscillator. Because the carrier frequency is around 2.4~GHz, direct software measurement is impractical. SC$\mu$M therefore provides an LC counter that counts a divided version of the carrier frequency, with a division ratio of 960. By reading and resetting this counter once per calibration window, LightCal can track carrier-frequency changes indirectly.

In LightCal, the LC counter is read and reset once every 12 synchronization pulses. When the RF carrier is correctly tuned, the accumulated LC count over a 100~ms window should be 250{,}000. We denote this target value by $LC_{\mathrm{ideal}}$. By comparing the measured count against $LC_{\mathrm{ideal}}$, LightCal estimates the carrier-frequency offset and updates the tuning parameters accordingly.

The LC oscillator is controlled by three digital parameters: \texttt{coarse}, \texttt{mid}, and \texttt{fine}, with the tuning relation
\[
1~\texttt{coarse} = 32~\texttt{mid} = 1024~\texttt{fine}.
\]
Starting from an initial setting, LightCal updates these parameters step by step and reevaluates the LC count after each calibration window. Once the residual error falls below a preset threshold, the current setting is considered sufficiently accurate for the next stage.

\subsection{Practical Robustness Under Polling Errors}

Although accumulation improves the effective timing stability of the optical reference, LightCal is still constrained by the SC$\mu$M platform. In Lighthouse V1, a synchronization pulse is only tens of microseconds longer than a sweep pulse. Under software polling, this margin is small enough that some synchronization pulses may be misclassified as sweep pulses, leading to missed detections and incorrect calibration-window boundaries.

To improve robustness, we introduce a differential filtering method. The key observation is that the LC frequency of SC$\mu$M is stable enough that the effect of a one-step tuning update is near-constant \cite{luo2024inter}. For example, increasing \texttt{mid} by one changes the LC count over a 100~ms window by approximately 80 ticks.
Therefore, when two consecutive calibration measurements correspond to adjacent valid tuning states, the difference between their LC counts should be close to a predictable constant value.

Let the LC counter values measured in two consecutive valid calibration windows be \(LC_1\) and \(LC_2\). We define
\begin{equation}
\Delta LC = \left| LC_2 - LC_1 \right|.
\end{equation}
If both windows are based on correct 12-sync accumulation, then \(\Delta LC\) should be close to an expected value \(\Delta LC_{\mathrm{exp}}\). We therefore accept the measurement pair only when
\begin{equation}
\left| \Delta LC - \Delta LC_{\mathrm{exp}} \right| \leq T_{\Delta LC},
\end{equation}
where \(T_{\Delta LC}\) is a small allowable deviation threshold.

In practice, LightCal records the LC count from one calibration step, compares it with the count from the next step, and uses the difference to determine whether the measurements are reliable. If the observed difference is close to \(\Delta LC_{\mathrm{exp}}\), the calibration update is accepted. Otherwise, the current measurement is discarded and the next window is reevaluated. This differential check helps suppress errors caused by polling-induced pulse misclassification without adding significant node-side complexity.

The main pipeline of Differential Filter is stated in \textbf{Algorithm \ref{alg:diff_filter}}.

\subsection{Coarse Optical Calibration with Fine Sweeping}

Differential filtering improves robustness, but it does not eliminate the fact that a substantial fraction of software-polled measurements still falls outside BLE-level accuracy. To ensure successful packet transmission after bootstrap calibration, LightCal uses the optical reference to reduce the carrier error to a bounded residual range and then resolves the remaining uncertainty through transmit-time fine sweeping.

Specifically, LightCal does not attempt to determine the \texttt{fine} parameter optically. Instead, it calibrates only \texttt{coarse} and \texttt{mid}. The residual uncertainty is then bounded by the allowable range of a single \texttt{mid} update. The corresponding relative error can be written as
\begin{equation}
\text{Relative Error (ppm)} =
\frac{\Delta \text{LC Counter}}{LC_{\mathrm{ideal}}}\times 10^6,
\label{eq:mid_error}
\end{equation}
where \(\Delta \text{LC Counter} = \pm 50\) and \(LC_{\mathrm{ideal}} = 250{,}000\). This yields a maximum residual error of about 200~ppm, which is much easier to satisfy reliably under polling-induced uncertainty than the full BLE target.

After \texttt{coarse} and \texttt{mid} are determined, the transmitter sweeps all 32 \texttt{fine} values under the selected \texttt{mid} setting during packet transmission. This hybrid design significantly relaxes the precision requirement on the optical reference while still ensuring that at least one transmitted packet is close enough to the target BLE channel frequency for practical reception.

\begin{figure}[t]
    \centering
    \includegraphics[width=0.78\linewidth]{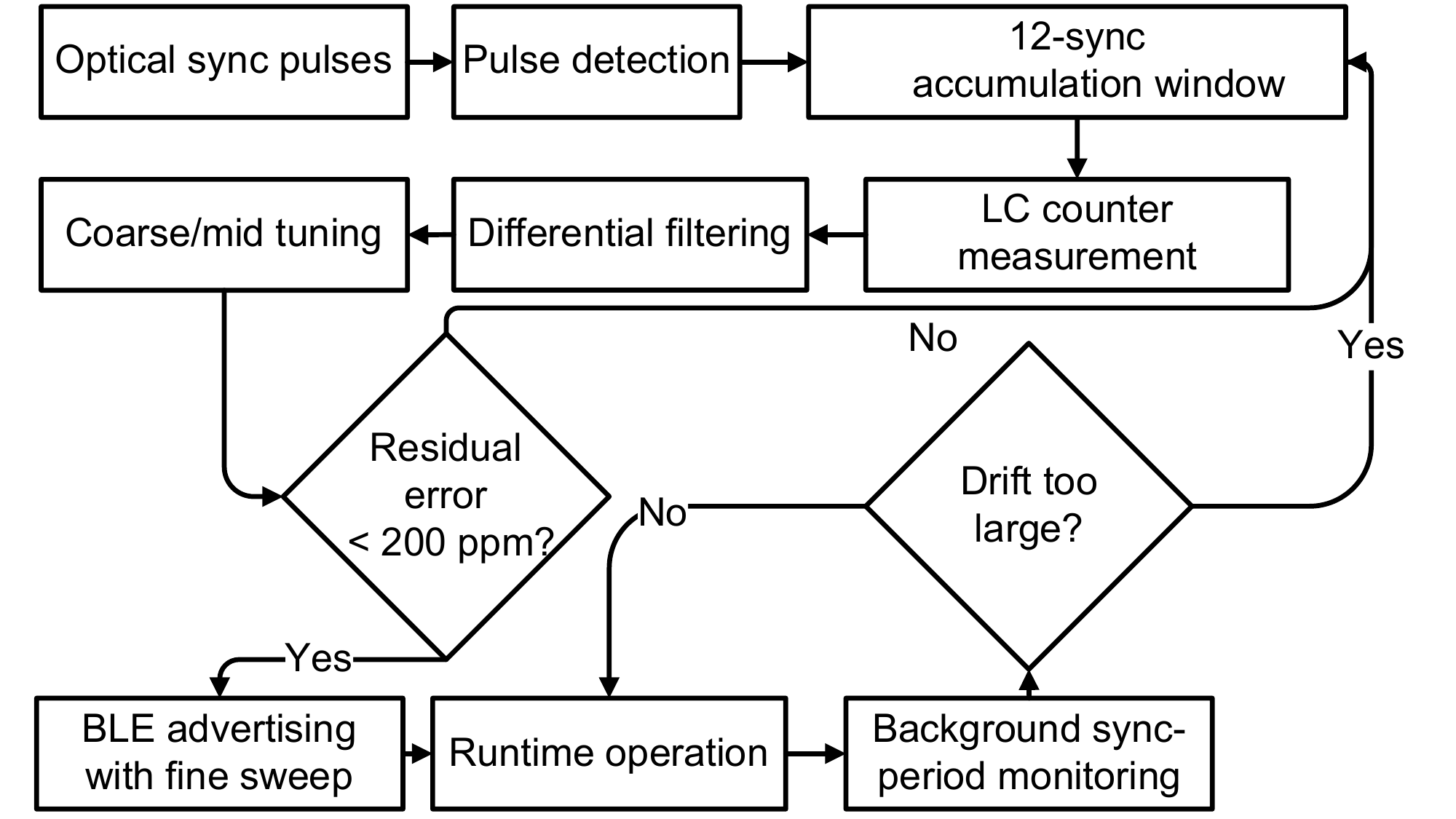}
    \caption{The \textbf{LightCal} workflow.}
    \label{fig:workflow}
\end{figure}

The complete accumulation, filtering, and sweeping process of LightCal is summarized in Fig.~\ref{fig:workflow}.
\vspace*{1mm}

\vspace*{2mm}
\begin{algorithm}[tb]
\caption{\textbf{Differentially filtered calibration update}}
\label{alg:diff_filter}
\renewcommand{\algorithmicrequire}{\textbf{Input:}}
\renewcommand{\algorithmicensure}{\textbf{Output:}}
\begin{algorithmic}[1]
\REQUIRE Current tuning state $(coarse, mid)$; expected LC-count difference $\Delta LC_{\mathrm{exp}}$; thresholds $T_{LC}$ and $T_{\Delta LC}$
\ENSURE Updated tuning state $(coarse, mid)$ and a validity flag

\STATE $LC_{\mathrm{meas}} \gets$ CountLCoverSyncWindow$(W)$
\IF{$|LC_{\mathrm{meas}} - LC_{\mathrm{ideal}}| \leq T_{LC}$}
    \STATE $(coarse, mid) \gets$ UpdateStep\\$(coarse, mid, LC_{\mathrm{meas}}, LC_{\mathrm{ideal}})$
    \STATE \textbf{return} $(coarse, mid, \textit{valid})$
\ENDIF

\IF{no previous valid measurement exists}
    \STATE store $LC_{\mathrm{meas}}$ as $LC_{\mathrm{prev}}$
    \STATE \textbf{return} $(coarse, mid, \textit{invalid})$
\ENDIF

\STATE $\Delta LC \gets |LC_{\mathrm{meas}} - LC_{\mathrm{prev}}|$
\IF{$|\Delta LC - \Delta LC_{\mathrm{exp}}| \leq T_{\Delta LC}$}
    \STATE $(coarse, mid) \gets$ UpdateStep\\$(coarse, mid, LC_{\mathrm{meas}}, LC_{\mathrm{ideal}})$
    \STATE $LC_{\mathrm{prev}} \gets LC_{\mathrm{meas}}$
    \STATE \textbf{return} $(coarse, mid, \textit{valid})$
\ELSE
    \STATE discard $LC_{\mathrm{meas}}$
    \STATE \textbf{return} $(coarse, mid, \textit{invalid})$
\ENDIF
\end{algorithmic}
\end{algorithm}

\section{Evaluation}

In this section, we evaluate LightCal from three perspectives. First, we compare its calibration behavior with a wired reference to understand whether optical triggering can support stable iterative tuning. Second, we examine whether the resulting calibration is sufficient for practical BLE advertising. Third, we test whether LightCal can coexist with Lighthouse-based localization during continued optical monitoring.

\subsection{Comparison with Wired Reference}

\begin{figure}[t]
    \centering
    \includegraphics[width=0.8\linewidth]{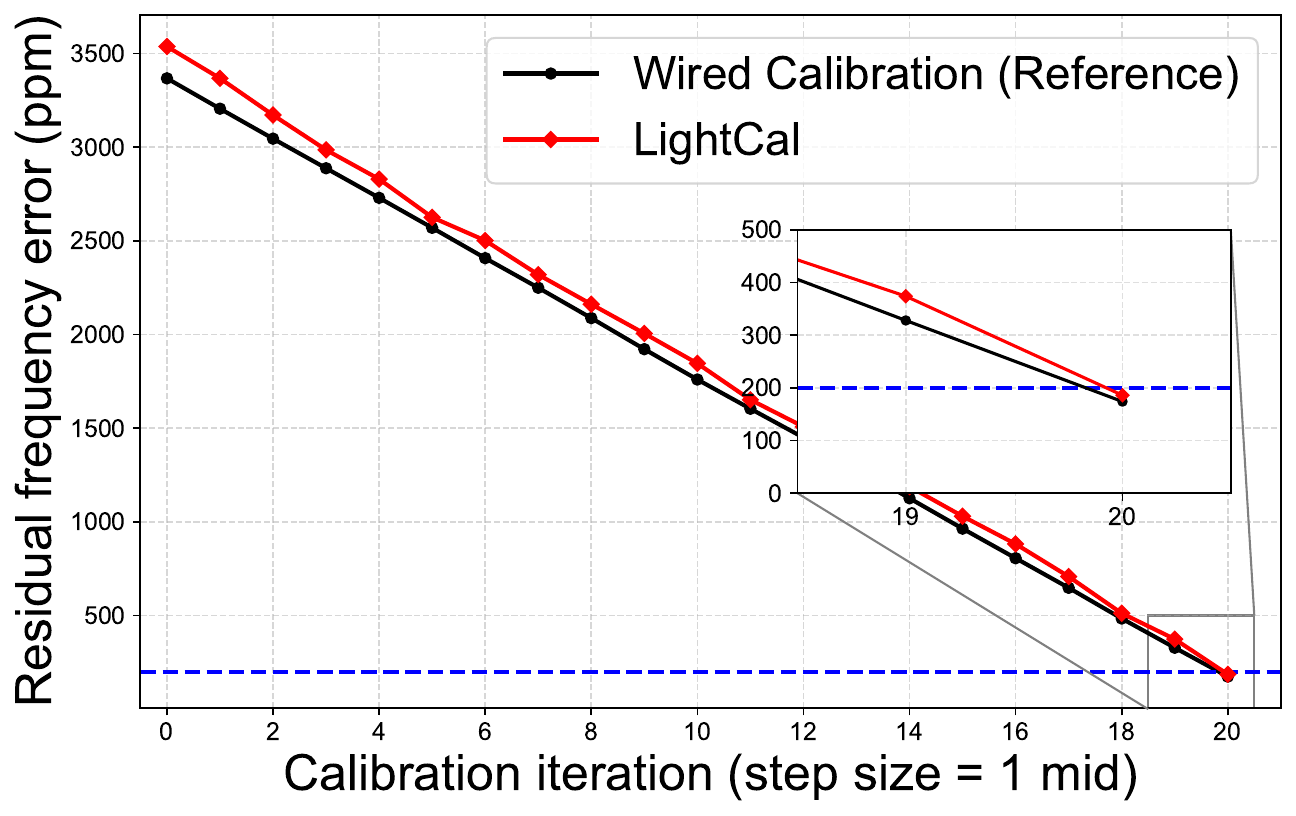}
    \caption{Calibration performance comparison between LightCal and the wired reference.
    }
    \label{fig:reference_compare}
\end{figure}

To compare LightCal with a wired triggered reference, we configure both methods to use the same tuning step size and time window. In the wired setup, an nRF52840 programmer is used to trigger the optical receiver of SCµM at 100 ms intervals during startup. The programmer can transmit a special pulse sequence that invokes a hardware-level interrupt, making the resulting calibration relatively accurate and suitable as a reference.

Fig.~\ref{fig:reference_compare} plots residual frequency error versus calibration iteration. The results show that LightCal, like the wired reference, steadily reduces the carrier-frequency error over successive iterations. In the current SC$\mu$M prototype, both methods reduce the residual error to approximately 100 LC counts, after which the remaining offset is handled by fine sweeping during transmission.

Compared with the wired reference, the LightCal trajectory exhibits modest step-to-step fluctuations, which are consistent with polling-induced uncertainty in sync-pulse detection. Even so, its convergence trend remains close to that of the wired reference throughout the calibration process. These results indicate that the optical reference is sufficiently stable to support practical iterative bootstrap tuning on SC$\mu$M.

\subsection{BLE Packet Visibility After Calibration}

\begin{figure}[!t]
	\centering
  \begin{subfigure}[t]{0.6\linewidth}
    \centering
    \includegraphics[width=\linewidth]{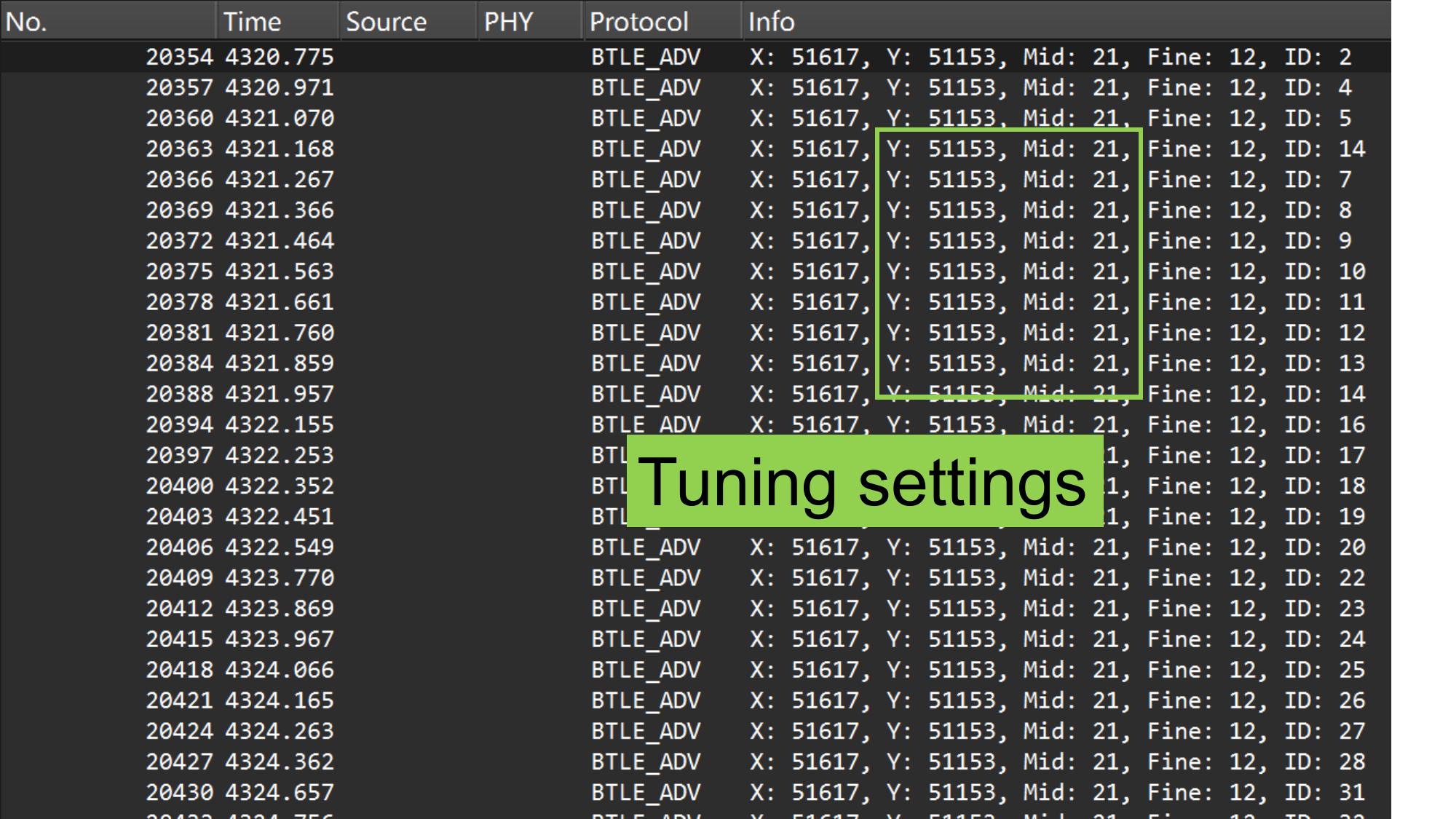}
    \caption{Example advertisement data extracted by wireshark.}
    \label{fig:pdr_extract_packet}
  \end{subfigure}
  \begin{subfigure}[t]{0.6\linewidth}
    \centering
    \includegraphics[width=\linewidth]{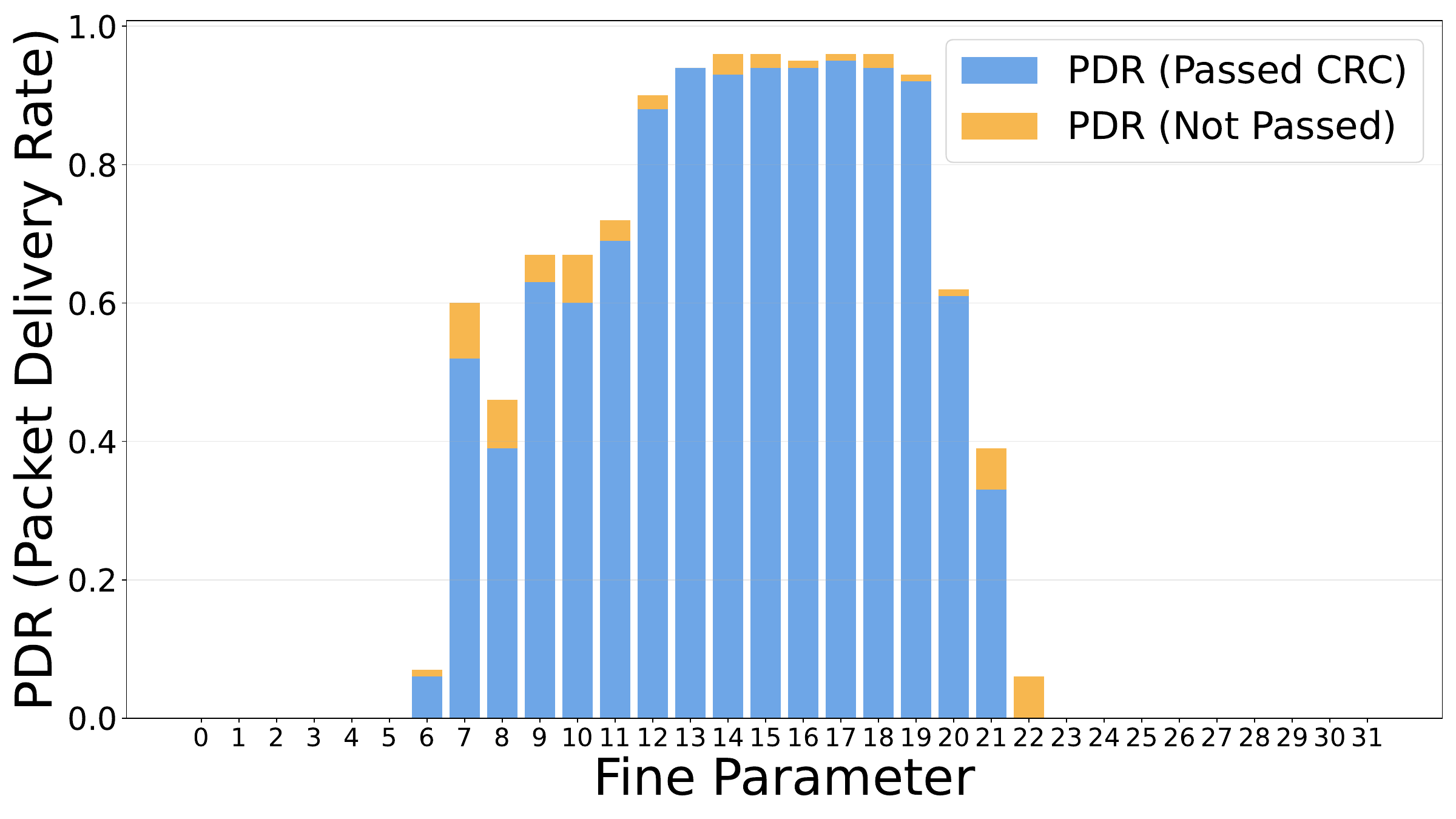}
    \caption{Packet delivery ratio across the 32 \texttt{fine} values within one \texttt{mid} setting, measured with an nRF sniffer.}
    \label{fig:pdr_result}
  \end{subfigure}
  
	\caption{
        BLE advertising visibility after LightCal calibration.
    }
	\label{fig:pdr}
\end{figure}

Because of hardware limitations in SC$\mu$M, LightCal reduces the carrier-frequency error to within approximately 200~ppm and then relies on fine sweeping during transmission. To verify that this level of calibration is sufficient for practical communication, we configure SC$\mu$M to transmit BLE advertising packets on Channel~37. Each packet carries basic information together with the current \texttt{mid}/\texttt{fine} settings. On the receiver side, we capture advertisements using an nRF sniffer and Wireshark on a PC (Fig.~\ref{fig:pdr_extract_packet}), and we classify packet reception by whether the captured packet passes the CRC check.

Fig.~\ref{fig:pdr_result} shows that receivable packets cluster near the center of the swept \texttt{fine} range within one \texttt{mid} span. This indicates that the target Channel~37 frequency of 2.402~GHz lies within the frequency region selected by LightCal. As the transmitted frequency approaches the center of this region, both the packet delivery ratio (PDR) and the fraction of CRC-valid packets increase. These results show that LightCal can bring the SC$\mu$M transmitter sufficiently close to the BLE channel center frequency for practical packet reception, with the remaining residual offset resolved by the final fine sweep.

\subsection{Compatibility with Lighthouse-Based Localization During Continued Optical Monitoring}

Although LightCal is designed for bootstrap radio calibration rather than localization, it continues monitoring Lighthouse synchronization pulses after the initial calibration step. This is necessary because the SC$\mu$M RF frequency may drift over time due to temperature or other environmental changes. Therefore, after the radio has been brought close enough to the BLE channel for transmission, LightCal keeps measuring sync-pulse timing in the background to detect whether recalibration is needed.

To evaluate whether this continued optical monitoring interferes with Lighthouse-based localization, we test a repeated localization--broadcast loop. In each cycle, SC$\mu$M performs Lighthouse-based localization while keeping the sync-period measurement path active for drift monitoring. In our short-term experiment, the measured sync period remains stable enough that no additional LC recalibration is triggered, although pulse timing continues to be collected throughout the run.

\begin{figure*}[t]
  \centering

  \begin{subfigure}[t]{0.22\linewidth}
    \centering
    \includegraphics[width=\linewidth]{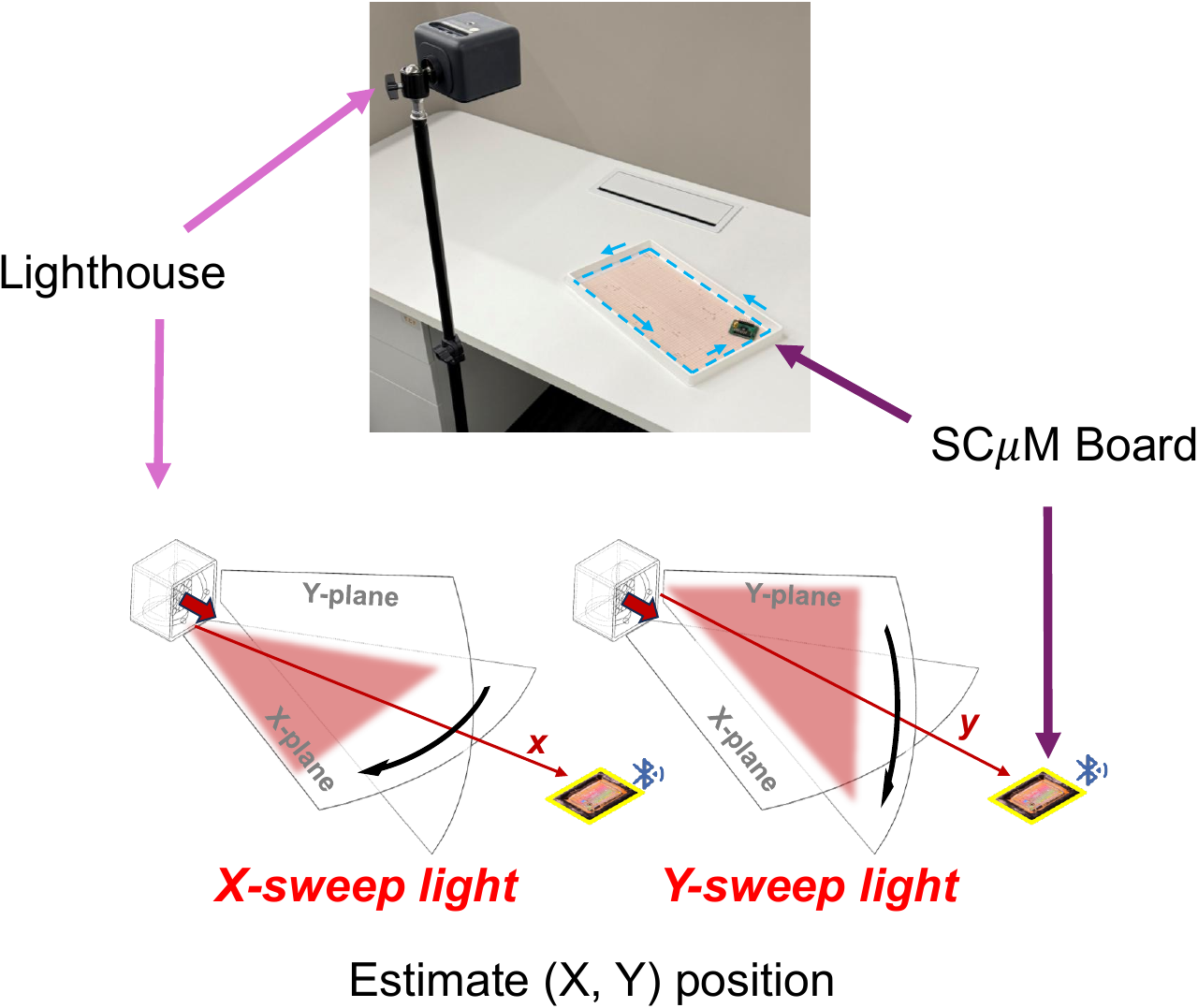}
    \caption{Experimental setup \\and localization principle.}
    \label{fig:eva_track_setup}
  \end{subfigure}
      \hfill
  \begin{subfigure}[t]{0.18\linewidth}
    \centering
    \includegraphics[width=\linewidth]{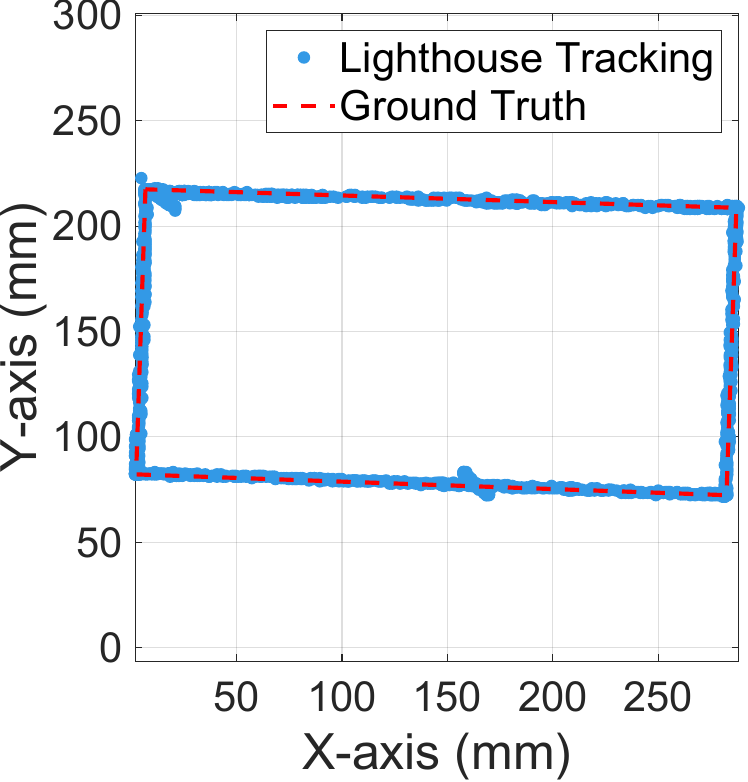}
    \caption{Reconstructed trajectory and ground truth.}
    \label{fig:eva_track_trace}
  \end{subfigure}
  \hfill
    \begin{subfigure}[t]{0.19\linewidth}
    \centering
    \includegraphics[width=\linewidth]{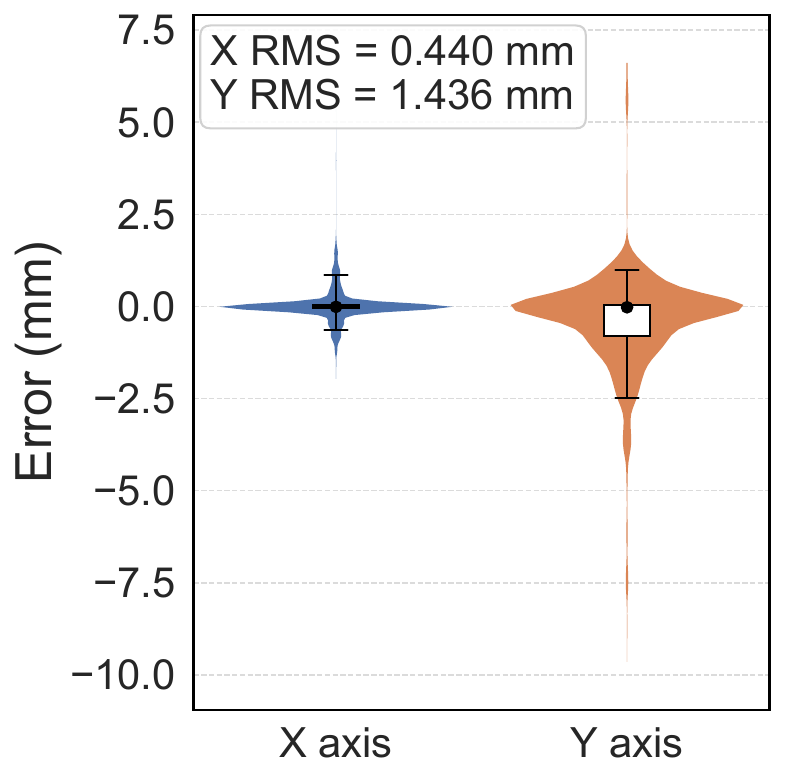}
    \caption{Tracking error.}
    \label{fig:eva_track_error}
  \end{subfigure}
      \hfill
  \begin{subfigure}[t]{0.2\linewidth}
    \centering
    \includegraphics[width=\linewidth]{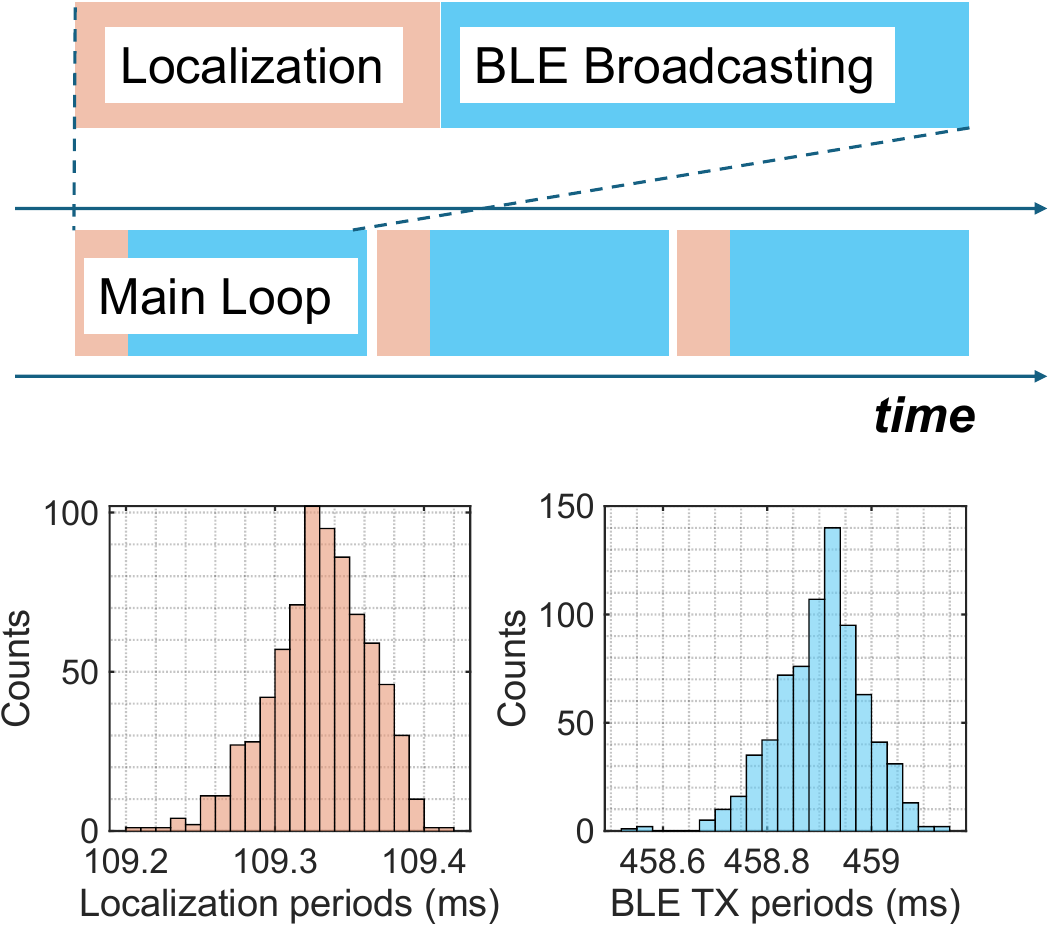}
    \caption{Localization-and-broadcast loop latency.}
    \label{fig:eva_track_latency}
  \end{subfigure}

  \caption{Compatibility of LightCal with Lighthouse-based localization.
  }
  \label{fig:eva_localization}
\end{figure*}

As shown in Fig.~\ref{fig:eva_track_setup}, SC$\mu$M is placed on a tabletop and moved along the edge of a box. 
The reconstructed two-dimensional trajectory in Fig.~\ref{fig:eva_track_trace} shows a clear rectangular path that closely follows the ground-truth motion. To further quantify localization quality, Fig.~\ref{fig:eva_track_error} plots the tracking error along the \(x\)- and \(y\)-axes. The measured errors in both directions remain predominantly at the millimeter level, indicating that keeping LightCal's pulse-monitoring active does not noticeably degrade Lighthouse-based localization accuracy in this experiment.

We also record firmware timestamps and append them to the BLE payload to quantify loop latency, as shown in Fig.~\ref{fig:eva_track_latency}. Each localization-and-broadcast cycle takes about 560~ms in total, of which about 460~ms is spent on fine sweeping during BLE transmission. Thus, the dominant delay comes from the transmit-side frequency sweep rather than from any conflict between localization and LightCal. Overall, these results show that Lighthouse-based localization and LightCal's continued pulse monitoring can coexist on SC$\mu$M without functional interference, while preserving millimeter-level localization accuracy in the tested setting.

\section{Discussion and Future Work}

The current LightCal prototype is mainly limited by the SC$\mu$M platform itself. 
Specifically, the current version of SC$\mu$M does not expose a usable interrupt interface for the optical receiver, so pulse timing must be obtained through software polling. This polling-only design is the main reason why the current implementation requires accumulation, differential filtering, and a final transmit-time fine sweep. 
We expect a future SC$\mu$M version that can provide hardware interrupts for optical events, so the timing accuracy of LightCal could be improved substantially, potentially eliminating the need for fine sweeping. LightCal also depends on line of sight and a suitable pulse source; future work will evaluate dedicated optical beacons and robustness under varying lighting conditions.

\section{Conclusion}

This paper presents \textit{LightCal}, a lightweight optical-pulse bootstrap calibration method for crystal-free BLE radios. By using periodic optical pulses as an external timing reference, LightCal addresses a key pre-RF deployment challenge for crystal-free nodes: obtaining a usable carrier-frequency reference before any network-assisted calibration becomes available. Our prototype implementation on SC$\mu$M with an unmodified commercial Lighthouse V1 source shows that low-complexity optical timing can support practical BLE bootstrap calibration using only pulse-window counting, LC-counter comparison, and lightweight filtering logic. Although the current prototype remains limited by polling-based optical capture and therefore still relies on a final fine sweep, the results demonstrate the practicality of optical references as a one-to-many bootstrap mechanism for crystal-free and highly integrated IoT platforms.

\section{Acknowledgment}
This work is funded by 
the Guangdong Provincial Key Lab of ICSC-IOT (2023B1212010007), 
the Guangzhou Municipal Science and Technology Project (2023A03J0011), 
the Guangzhou Municipal Key Laboratory on Future Networked Systems (2024A03J0623) and 
the Guangdong Provincial ZJ talent scheme 2025D03J0007.

\bibliographystyle{IEEEtran}
\bibliography{wang25loca}

@inproceedings{wheeler2019low,
  title={A low-power optical receiver for contact-free programming and 3D localization of autonomous microsystems},
  author={Wheeler, Brad and Ng, Andrew and Kilberg, Brian and Maksimovic, Filip and Pister, Kristofer SJ},
  booktitle={2019 IEEE 10th Annual Ubiquitous Computing, Electronics \& Mobile Communication Conference (UEMCON)},
  pages={0371--0376},
  year={2019},
  organization={IEEE}
}

@article{kilberg2020accurate,
  title={Accurate 3D lighthouse localization of a low-power crystal-free single-chip mote},
  author={Kilberg, Brian G and Campos, Felipe Mulinari Rocha and Maksimovic, Filip and Watteyne, Thomas and Pister, Kristofer SJ},
  journal={Journal of Microelectromechanical Systems},
  volume={29},
  number={5},
  pages={818--824},
  year={2020},
  publisher={IEEE}
}

@inproceedings{maksimovic2019crystal,
  title={A crystal-free single-chip micro mote with integrated 802.15. 4 compatible transceiver, sub-mw ble compatible beacon transmitter, and cortex m0},
  author={Maksimovic, Filip and Wheeler, Brad and Burnett, David C and Khan, Osama and Mesri, Sahar and Suciu, Ioana and Lee, Lydia and Moreno, Alex and Sundararajan, Arvind and Zhou, Bob and others},
  booktitle={2019 Symposium on VLSI Circuits},
  pages={C88--C89},
  year={2019},
  organization={IEEE}
}

@inproceedings{alghaihab202030,
  title={30.7 A Crystal-Less BLE Transmitter with- 86dBm Freq $\mu$ ency-Hopping Back-Channel WRX and Over-the-Air Clock Recovery from a GFSK-Modulated BLE Packet},
  author={Alghaihab, Abdullah and Chen, Xing and Shi, Yao and Truesdell, Daniel S and Calhoun, Benton H and Wentzloff, David D},
  booktitle={2020 IEEE International Solid-State Circuits Conference-(ISSCC)},
  pages={472--474},
  year={2020},
  organization={IEEE}
}

@inproceedings{wang2024simultaneous,
  title={Simultaneous Localization and Clock Calibration for Crystal-Free Mote},
  author={Wang, Cheng and Chang, Tengfei and Alvarado-Marin, Said and Burnett, David and Maksimovic, Filip and Watteyne, Thomas and Pister, Kristofer SJ},
  booktitle={2024 IEEE Workshop on Crystal-Free/-Less Radio and System-Based Research for IoT (CrystalFreeIoT)},
  pages={36--37},
  year={2024},
  organization={IEEE}
}

@inproceedings{luo2024inter,
  title={Inter-Cal: Inter-Oscillator Calibration for Crystal-Free Mote-on-Chip},
  author={Luo, Yuanming and Chang, Tengfei and Burnett, David and Maksimovic, Filip and Watteyne, Thomas and Pister, Kristofer SJ and He, Jie},
  booktitle={2024 IEEE Workshop on Crystal-Free/-Less Radio and System-Based Research for IoT (CrystalFreeIoT)},
  pages={12--17},
  year={2024},
  organization={IEEE}
}

@article{chang2021surviving,
  title={Surviving the Hair Dryer: Continuous Calibration of a Crystal-Free Mote-on-Chip},
  author={Chang, Tengfei and Watteyne, Thomas and Wheeler, Brad and Maksimovic, Filip and Burnett, David C and Pister, Kris},
  journal={IEEE Internet of Things Journal},
  volume={9},
  number={6},
  pages={4737--4747},
  year={2021},
  publisher={IEEE}
}

@article{chang2020quickcal,
  title={Quickcal: Assisted calibration for crystal-free micromotes},
  author={Chang, Tengfei and Watteyne, Thomas and Maksimovic, Filip and Wheeler, Brad and Burnett, David C and Yuan, Titan and Vilajosana, Xavier and Pister, Kristofer SJ},
  journal={IEEE internet of things journal},
  volume={8},
  number={3},
  pages={1846--1858},
  year={2020},
  publisher={IEEE}
}

@inproceedings{wheeler2017crystal,
  title={Crystal-free narrow-band radios for low-cost IoT},
  author={Wheeler, Brad and Maksimovic, Filip and Baniasadi, Nima and Mesri, Sahar and Khan, Osama and Burnett, David and Niknejad, Ali and Pister, Kris},
  booktitle={2017 IEEE Radio Frequency Integrated Circuits Symposium (RFIC)},
  pages={228--231},
  year={2017},
  organization={IEEE}
}

@article{stanislowski2013adaptive,
  title={Adaptive synchronization in IEEE802. 15.4 e networks},
  author={Stanislowski, David and Vilajosana, Xavier and Wang, Qin and Watteyne, Thomas and Pister, Kristofer SJ},
  journal={IEEE Transactions on Industrial Informatics},
  volume={10},
  number={1},
  pages={795--802},
  year={2013},
  publisher={IEEE}
}

@inproceedings{yuan2026automatic,
  title={Automatic Network-Based Multi-Channel Frequency Calibration for Self-Joining Crystal-Free Motes},
  author={Yuan, Titan and Maksimovic, Filip and Chang, Tengfei and Pister, Kristofer S. J.},
  booktitle={Proceedings of the 2026 International Conference on Embedded Wireless Systems and Networks (EWSN)},
  year={2026},
  note={Accepted for publication}
}

@inproceedings{song202030,
  title={30.8 A 3.5 mm$\times$ 3.8 mm crystal-less MICS transceiver featuring coverages of$\pm$160ppm carrier frequency offset and 4.8-VSWR antenna impedance for insertable smart pills},
  author={Song, Minyoung and Ding, Ming and Tiurin, Evgenii and Xu, Kai and Allebes, Erwin and Singh, Gaurav and Zhang, Peng and Traferro, Stefano and Korpela, Hannu and Van Helleputte, Nick and others},
  booktitle={2020 IEEE International Solid-State Circuits Conference-(ISSCC)},
  pages={474--476},
  year={2020},
  organization={IEEE}
}

@article{xu2016energy,
  title={Energy-efficient time synchronization in wireless sensor networks via temperature-aware compensation},
  author={Xu, Miao and Xu, Wenyuan and Han, Tingrui and Lin, Zhiyun},
  journal={ACM Transactions on Sensor Networks (TOSN)},
  volume={12},
  number={2},
  pages={1--29},
  year={2016},
  publisher={ACM New York, NY, USA}
}

@inproceedings{ding20200,
  title={A 0.9 pJ/cycle 8ppm/° C DFLL-based wakeup timer enabled by a time-domain trimming and an embedded temperature sensing},
  author={Ding, Ming and Song, Minyoung and Tiurin, Evgenii and Traferro, Stefano and Liu, Yao-Hong and Bachmann, Christian},
  booktitle={2020 IEEE Symposium on VLSI Circuits},
  pages={1--2},
  year={2020},
  organization={IEEE}
}

@article{movassaghi2014wireless,
  title={Wireless body area networks: A survey},
  author={Movassaghi, Samaneh and Abolhasan, Mehran and Lipman, Justin and Smith, David and Jamalipour, Abbas},
  journal={IEEE Communications surveys \& tutorials},
  volume={16},
  number={3},
  pages={1658--1686},
  year={2014},
  publisher={IEEE}
}


\end{document}